\documentstyle[12pt]{article}
\begin{document}

\begin{center}
{\large \bf Self-Microlensing in Compact Binary Systems}
  \footnote[1]{Supported by the National  Science 
   Foundation of China under Grant No. 19421004}\\[5mm]

{ QIN Bo, \ \  WU Xiang-Ping, \ \  ZOU Zhenlong}\\ 
{Beijing Astronomical Observatory, Chinese Academy of Sciences,
       Beijing 100080}\\[10mm]

{\bf ABSTRACT}\\
\end{center}

The signature of the self-microlensing in compact binaries 
(white dwarfs, neutron stars and black holes) is a flare
with the characteristic time of typically a few minutes. 
The probability of detecting these microlensing events 
can be as high as $1/50$ for a flux sensitivity of $\Delta m=0.01$ in
magnitude. The discovery of the self-microlensing by binaries would
furnish an additional way to find the masses of the lens and 
the companion and will be promising for the searches of black holes.\\

PACS: 95.30.Sf, 97.80.-d, 97.60.Jd\\

\newpage

The detection of the gravitational microlensing events and their 
relatively low detection rate  
in the Large Magellanic cloud (LMC) $^{1}$ have stimulated 
several studies on the possibility of
attributing the LMC microlensing events to the stars and/or 
MAssive Compact Halo Objects (MACHOs)
of the LMC itself $^{2,3}$.  Gould $^{4}$  referred to such a
phenomenon  as ``self-lensing'' and extended this research to 
the binary systems $^{5}$. 
On the other hand, the binary stars
are numerous and some of the microlensing events seen in the 
Galactic bulge are actually due to the microlensing 
by ordinary star binary systems $^{6}$. Here we use the term
``compact binary'' to denote a binary system composed of two compact objects.\\

Gould $^{5}$ has estimated the optical depth $\tau$ to self-microlensing 
in various binary systems. It turns out that for a star
binary system $\tau$ is too
small ($\sim10^{-11}$) to have any observational significance, while
compact binaries composed of neutron stars or black
holes may have a somewhat larger $\tau$ of up to $0.2\%$ 
to generate  observational features. In terms of these estimates,
it appears to be very difficult and even hopeless to 
search for self-microlensing events in binaries considering the fact that 
the present-day discovered pulsar binaries are only a few tens. \\

We notice, however, that the optical depth to gravitational lensing
only accounts for those events whose positions happen to appear within the
Einstein rings of the lensing objects, which corresponds to a 
magnification of $\mu=1.34$ or an apparent magnitude of $\Delta m=0.32$.
While the advanced detectors in today's astronomical observations
can have a sensitivity of as high as $\Delta m\sim10^{-3}$, the lensing
probability $p$ can be greatly increased if the small magnification events
are taken into account.  For instance, the above claimed optical
depth of binary systems can be raised by a factor of $\sim10$ if
one includes the $\Delta m>0.01$ events. This change is probably trivial for
ordinary star binaries due to their extremely low optical depth, but
would be of great significance for the compact binary systems. Actually,
it will be promising to detect the self-lensing in a binary system
composed of two compact objects like millisecond pulsars whose lensing 
probability is $p\sim2\%$ for a sensitivity of $\Delta m=0.01$.
Motivated by this large lensing probability ($\sim1/50$) for self-lensing 
in compact binary systems and the $\sim10$ discovered 
pulsar/black hole binaries among the 706 pulsars observed so far, 
this paper presents an  analysis of
the microlensing features arising from the self-lensing in 
compact binaries and a tentative search for the 
self-lensing signatures from the existing catalogs. \\

We begin with a binary system composed of two compact objects 
with identical mass $M$ and
neglect their sizes and the possible occulation. Furthermore, 
we assume their rotating orbit to be a circle with separation of
$r$. If we utilize $D_d$ and $D_s$ to denote 
the angular diameter distances from the observer to
the lens and to the source, respectively,
we have  $D_d\approx D_s$. Either of the compact 
objects is able to gravitationally magnify the brightness of its 
companion, depending on their positions and the orbital inclination
$i$. The combined magnification of the lens-induced two images is
\begin{equation}
\mu=\mu_1+\mu_2=\frac{\ell^2+2a_E^2}{\ell\sqrt{\ell^2+4a_E^2}},
\end{equation}
where $\ell$ is the alignment parameter or the true position 
of the lensed companion
\begin{equation}
\ell=r\sqrt{1-\sin^2i \cos^2\theta},
\end{equation}
$a_E$ is the Einstein radius
\begin{equation}
a_E=\sqrt{\frac{4GM}{c^2}r\sin i\cos\theta}
\end{equation}
and $\theta$ measures the position of the lensing object along its orbit with
$\theta=0$ at the closest point to the observer, i.e.,  $\mu$ takes
the maximum at $\theta=0$. The orbital motion of the source and lens would
lead to the variation of the source brightness. Figure 1 shows a typical
variation of source magnification  within an orbital period ($P$)
for a nearly edge-on compact binary composed of two objects with
mass of $1M_{\odot}$ and separation of $R_{\odot}$. 
The lensing object temporarily magnifies its companion when the alignment
parameter of the companion object reaches the minimum, which leads to a
flare on the brightness of the companion. The characteristic time
of this flare can be described by 
\begin{equation}
\Delta T=\frac{P}{\pi}\sqrt{\frac{4GM}{c^2r}u(\mu_{min})
-\left(\frac{\pi}{2}-i\right)^2},
\end{equation}
where $\mu_{min}$ is the minimum magnification corresponding to the 
flux sensitivity of the observation. It appears 
that for the example of Fig.1 and a period of $P=1$ day, the
time resolution should reach $\Delta T=5 (2)$ minutes to detect
the $\Delta m>0.01(0.1)$ event. Yet, it turns out to be feasible in
observations, though it is much shorter than the timescale of the
microlensing events in the LMC and in the Galactic bulge. 
Two flares in the combined light curves would occur within a period 
when both objects have electromagnetic radiation. \\

Unlike the microlensing events in the LMC and the galactic bulge
where the distance and the transverse velocity of the lens
are two unknown parameters, the microlensing in binaries is well
constrained by their orbital parameters ($i$, $P$ and $r$). This 
furnishes an additional way to estimate the mass of the lens using only
the maximum magnification $\mu_{max}$
\begin{equation}
M=\frac{c^2r}{8G}(\sin^{-1} i-\sin i)
  \sqrt{\mu_{max}^2-1}(\mu_{max}+\sqrt{\mu_{max}^2-1}).
\end{equation}
In particular, if the inclination $i$ is close to $\pi/2$, we have
\begin{equation}
M=18M_{\odot}\sqrt{\mu_{max}^2-1}
  (\mu_{max}^2+\sqrt{\mu_{max}^2-1})\;
  \left(\frac{r}{R_{\odot}}\right) 
  \left(\frac{90^{\circ}-i}{1^{\circ}}\right)^2.
\end{equation}
Once the mass of the lens is determined, one can easily found the 
mass of the companion according to the Keplerian law.  This method
would be of great interest to the searches for black holes in 
binary systems. \\

We now examine our working hypothesis, i.e.,
the binaries are assumed to be pointlike. The Einstein 
radius from eq.(3) is estimated to be 
\begin{equation}
a_E\sim2.9\times10^{-3}\;\left(\frac{r}{R_{\odot}}\right)^{1/2}
        \left(\frac{M}{M_{\odot}}\right)^{1/2}\;R_{\odot}.
\end{equation}
A detection sensitivity of $\Delta m=0.01$ corresponds to a 
circle with radius of $3.6a_E$ around the lens, which is the 
size of the Earth ($R_{\oplus}$) for $r=R_{\odot}$ and $M=M_{\odot}$. 
Therefore, ordinary stars and the earth-mass 
planets cannot produce self-lensing effect due to their sizes
being much larger than their Einstein radii.   
A white dwarf in a binary system can marginally act as 
a lens for its companion and a large separation
of $\sim1$ AU further reduces the occultation effect. 
On the other hand, the maximum magnification of a white dwarf with
a size of $R_{\oplus}$ lensed by its compact companion with a mass of
$M_{\odot}$ can reach $\mu_{max}=(1.2,9.4)$ for the  separation of 
$r=(R_{\odot},1$ AU). It turns out that the white dwarf binaries 
can be considered as a good type of system for searches of self-lensing
though a pointlike approximation might fail for a
dwarf being the target source. 
As for a compact binary composed of two pulsars or black holes,
the pointlike assumption should be  a reasonable approximation since
their sizes are much less than the Einstein radius given by Eq.(7).\\

The updated pulsar catalog contains 706 sources, among which 44 are binary
systems including neutron star+ordinary star, neutron star+white dwarf,
neutron star+planet, neutron star+black hole candidate and neutron
star+neutron star. This number decreases if we exclude the
systems having ordinary stars and/or planets. Furthermore, our study is
limited to a few compact binaries since only $1/10$ of the binaries have  
the orbital inclinations $i$  observationally determined or constrained.\\

About $40\%$ of the binaries are neutron star+white dwarf systems 
$^{7}$.
An interesting example is the millisecond binary pulsar PSR B1855+09
 $^{8}$,
which moves along a nearly circular orbit with a period of 12.3 days. 
The mass of the pulsar and of the companion white dwarf 
are $1.50M_{\odot}$ and $0.258M_{\odot}$, respectively. 
In particular, its orbital plane is nearly edge-on with 
$\sin i=0.9992^{+0.0004}_{-0.0007}$. 
This large inclination makes it an attractive system for searches of
the self-lensing signature. Unfortunately, the too large projected semimajor 
axis of $x=9.23$ light seconds, which yields $r\approx 27 R_{\odot}$ and
the inclination $i=87^{\circ}.7$, cannot result in any observational lensing 
features.\\

There are only four neutron star+neutron star
systems and several neutron star+black hole candidate
systems discovered to date. However, $\sim10^5$
such systems are expected to exist in our Galaxy $^{9}$.
About $2\%$ of these binaries will show the signature of self-lensing
and $0.2\%$ of them would show strong self-lensing $\mu>1.34$. 
Undoubtedly, 
this will provide a great number of samples for testing the 
general relativity and for discovering the black holes. \\

We are very grateful to Dr. Shude Mao for his stimulating discussion.\\



\noindent Figure Caption\\

\noindent {{\bf Fig.1.} The microlensing induced brightness flare in a binary
system composed of two compact objects with mass of $1M_{\odot}$
and separation of $R_{\odot}$. The orbital inclination is taken
to be $89^{\circ}.9$. $P$ denotes the period and $\Delta m$ is the
magnitude.

\end{document}